# Formation of lightning
## in terms of opinion dynamics in three dimension

Çağlar Tuncay
Department of Physics, Middle East Technical University
06531 Ankara, Turkey
caglart@metu.edu.tr

**Abstract:** Formation of a lightning within a cloud, between clouds and towards the earth is studied as an application of sociophysics. The three dimensional society is sky or cloud, which has members as electrically charged clouds (in sky) or patches (in cloud). Members interact with the neighboring ones and all are convinced to average their charges (opinion). Yet, big external drives (mass media) as winds and turbulences may load new charges or may force the present ones to accumulate temporally at a site. For a lightning towards the earth, similarly charged clouds in sky (patches carrying big charges in a cloud) are expected to come close to each other. In all, discharging process is nothing, but what is called lightning.

**Introduction:** From the time of Benjamin Franklin (1706-1790) on, many scientific investigations were made on lightning. In this contribution, which is (up to our knowledge) the first application of 3-dimensional opinion dynamics and to meteorology, we represent a cloud (sky) in our model by a cubic (I=NxNxN) matrix, where each entry (i) carries a time (t) dependent charge $Q_i(t)$. ([**1**], and references therein.] We may have macroscopic neutrality and conservation of total charge;

$$\sum_i^I Q_i(t) = 0 \quad . \tag{1}$$

In fact this condition is optional and any cloud may carry a net charge ($Q_{net}$).

These are well known, that the corresponding potential ($V_i$) of each site, with respect to infinity is defined as;

$$V_i(t) = Q_i(t)/C_i(t) \quad , \tag{2}$$

where $C_i(t)$ is the local electrical capacity, and effect of other charged patches are ignored. We have the total potential ($V^T_i$) at (i) (approximately) as;

$$V^T_i(t) = V_i(t) + \sum_{i \neq j}^I Q_i(t)/r_{i,j} \quad , \tag{3}$$

where $r_{i,j}$ is the Euclidean distance between lattice points (i) and (j).

Vectoral electric field ($\underline{E}$) and current density ($\underline{J}$) between the close entries (i) and (j) are given as

$$\underline{E}_i(t) = (\nabla V^T(t))_i \quad , \tag{4}$$

and

$$\underline{J}(t) = \sigma \underline{E}(t) \quad , \tag{5}$$

respectively. In Eq. (4), the gradient of V ($\nabla V^T(t)$) is calculated at (i), and in Eq. (4) $\sigma$ is the usual (local) conductivity. Moreover we have a magnetic field $\underline{B}(t) = \partial \underline{E}/\partial t$, which we ignore.

Under normal conditions, i.e., if the conductivity is sufficiently large, neighboring charges neutralize each other; $Q_i(t) \to Q_i(\infty)$, and $V^T_i(t) \to V^T_i(\infty)$ as $t \to \infty$. Whereas, if big turbulences exist, where the charged patches (clouds) move with respect to each other with high speeds, neutralizations may not be completed at a given time. In the mean time, big charge accumulations may take place occasionally, so that the magnitude of $\underline{E}(t)$ (Eq. (4)) may exceed the critical value (sparking threshold) of the cloud (air), and lightning may occur within a cloud, between clouds and towards the earth.

Our model is given in the following section; applications and results are displayed in next one. Last section is devoted for discussion and conclusion.

**Model:** We assume uniform and constant capacitance and conductivity between the clouds and within a cloud, Eqns. (2), and (5), respectively, and take each as unity; $C_i(t)=1$, independently of time and the sub index, and $\sigma(t)=1$, for all t. Moreover we follow a first nearest neighbor (nn) approximation for connectivity with unit binding strength between each nn.

*Initiation:* We charge each entry by random real numbers $\pm Q_i(0)$, where $-1 \leq Q_i(0) \leq 1$. Clearly, Eq. (1) is satisfied (within randomness) and unit for charge is irrelevant here.

*Evolution:* We apply the usual iterative interaction tours, and assume that only nn charges interact and average[2] with $Q_i$;

$$Q_i(t) = (Q_i(t-1) + \sum_j^{nn} Q_j(t-1))/(\rho+1) \quad , \tag{6}$$

where $\rho$ is the number of nn for (i), and $\rho=6$ if (i) is in bulk, $\rho=5$ if (i) is on the surface, $\rho=4$ if (i) is on the edge, and $\rho=3$ if (i) is at the corner of the cubic cloud.

For turbulences, we randomly exchange charges at sites, ($Q_i(t) \leftrightarrow Q_j(t)$), i.e. we shuffle the charges at each tour; and then let mutual interactions take place between the nn's for averaging in terms of short-range currents (no spark, no lightning)). One may repeat this process at each tour and observe fluctuations in $\underline{E}(t)$ within the cloud (between the clouds). And whenever E(t) exceeds in magnitude the critical value (sparking threshold) of the cloud ($E_{cloud}$), lightning may occur between the sites (i) and (j) in a cloud.

For a lightning towards the ground, $V^T_i(t)/H$ must exceed the critical value (sparking threshold) of the air ($E_{air}$), where H is the height of the cloud and $E_{air}$ is lower than the sunny day values due to the present wetness.

**Results:** In Figure 1. a.-d. we display the averaging process of charges (opinions)[2] taking place within a cloud (society) matrix, where the matrix is charged by random real numbers $\pm Q_i(0)$ initially, as described within the previous section. Fig. 1. a. displays charge evolution of (representative) three sites (i), (N=20). Fig. 1. b. displays $E^{max}_{nn}(t)$ and $Q^T(t)$, for N=20, where $E^{max}_{nn}(t)$ is the maximum value in magnitude of the electric field between the nn sites, and $Q^T(t)$ is the total charge ($\sum^I_i Q_i(t)$) within the matrix, both at a given time t. Please note that $Q^T(t)$ is crucial since it involves the information about the evolution of charges macroscopically, and $E^{max}_{nn}(t)$ is crucial since it controls the ignition of lightning. Fig. 1. c. is the histogram for initial distribution of charges (opinions) within the cloud (society), with N=9. Fig. 1. d. is the charge distribution at the $200^{th}$ tour (time, t) with N=9, ($Q^T(t \to \infty) = Q_{net}$).

We represent a wind (storm, turbulence) by two parameters; (random) duration for averaging process ($t_{wait}$) between two consecutive shuffles (($Q_i(t) \leftrightarrow Q_j(t)$)), and by number of sites interchanged per shuffle. Clearly, $t_{wait}$ is a measure for wind speed and it is inversely proportional to speed. The number of charged sites interchanged per shuffle represents turbulence due to wind, where turbulence may also be taken as proportional to wind speed.

Magnitude of the electric field between any neighboring sites carrying opposite unit charges, may be calculated (Eqns. (3) and (4)) to be in arbitrary units. We may utilize some multiples of this value as a unit for spark thresholds. So; $E_{cloud} = C$, and $E_{air} = A$.

Figure 2. a.-c. (N=10) displays effect of wind speed and turbulence on $E^{max}_{nn}(t)$, where $E^{max}_{nn}(t)$ is the maximum value in magnitude of the electric field between the nn sites. Fig. 2. a. is $E^{max}_{nn}(t)$ with $t_{wait} \leq 1000$, i.e., at the end of a time period of $t_{wait}$ equals to 1000 at maximum. Wind will turbinate the cloud randomly by 0% (thinnest line), 20%, and 50% (thickest line). Fig. 2. b. and Fig. 2. c. is as Fig. 2. a. with $t_{wait}$ equals to 100 at maximum and $t_{wait}$ equals to 10 at maximum, respectively.

We tried several more combinations of $t_{wait}$ and number of interchanged sites per shuffle in many runs and observed that, the present mechanism has a small probability to form a lightning. Yet, it may ionize the matter at the interfaces of nn sites in terms of glowing and thus charge may flow easily through these channels. It may be emphasized that, turbulence does not cause lightning to occur but it may be useful for it.

On the other hand, turbulence may bring similar charges side by side, and increase the local potential temporarily. It may also suddenly squeeze the volume (volume$_i(t)$) and lessen the local capacity ($C_i$, in Eq. (2)). It is well known, that the electrical capacity is proportional to its space dimension in terms of length, thickness, radius, etc. So, for a given $Q_i(t)$, $V^T_i(t)$ may abruptly increase during (adiabatic) pressure ($P_i(t)$) vortices: $V^T_i(t) \propto (C_i(t) \propto (volume_i(t))^{-1/3} \propto) (P_i(t))^{1/3}$. Hence, if pressure increases by a factor of eight say (i.e. to P=8 atmosphere), then the local potential and the maximum of electric field in magnitude increase by a factor of two, and thus the threshold value C (A) may be exceeded, and we have a lightning.

In Figure 3. a.-b. we display $E^{max}_{nn}(t)$ for various $t_{wait}$ (N=5, and pressure increases by a factor of eight at maximum, in all).

Whenever the pressure increment within a cloud occurs regionally, i.e. covering many sites, we may have longer, intra clouds lightning. And sometimes, similar charges may be accumulated at a wider region, and the opposite ones may be distributed within the cloud. Then, averaging process and sparking between neighboring sites cease, which may cause a lightning to occur towards the ground. In this case, we may have Eq. (3) with $A \leq V^T/H$. In Figure 4. we display a charge distribution where we have regional distributions of similar charges within a cloud or in the sky.

**Discussion and Conclusion:** The present approach may be considered as qualitative, where many approximations are performed. It is clear that, at any time t and for a volume charge density $\rho(\underline{r})$, $\nabla^2 V(\underline{r}) = \rho(\underline{r})$ must be solved and the effect of $\underline{B}(r,t)$ must be taken into account for more accurate results. Moreover the probability density functions for Q, V, and E might be utilized for more expressive figures, if the present computer facilities were suitable for it. We ignored the possible smaller lightings which might have occurred at earlier times than lightning occurred.

In case of turbulence, instead of shuffling, new charge configurations may be set at each tour. And, as another equivalent way, one may choose some sites randomly and load there new random charges $Q'_i$. And, it is clear that, big clouds are likely to have inter cloud lightning and small ones are that to have intra cloud ones, and that is why we varied our matrix size N.

As a final remark, it may be stated that, the effect of storm, which increases the local potential in terms of pressure increment may be considered as a form of leader (media) effect in sociophysics. [**3**, and references therein.]

**Acknowledgement**
The author is thankful to Dietrich Stauffer for his friendly discussions and corrections, and informing about the references [**1, 2**].

**FIGURES**

**Figure 1.**     **a**. Charge evolution of (representative) 3 sites (i) (N=20) in terms of averaging (Eqn. (6)), when there is no wind, where the cloud is charged by random real numbers $\pm Q_{ij}(0)$ initially, as described within the relevant text.
**b.** $E^{max}_{nn}(t)$ and $Q^T(t)$, for N=20, where $E^{max}_{nn}(t)$ is the maximum value in magnitude of the electric field between the nn sites, and $Q^T(t)$ is the total charge within the matrix, both at a given time t.
**c.** Initial distribution of charges (N=9).
**d.** Charge distribution at t=200 (N=9).

**Figure 2.**     **a.** $E^{max}_{nn}(t)$ for $t_{wait}$ = 1000 (N=10), where the number of sites interchanged is varied from zero (thinnest line) to 500 (=0.50x10x10x10) (thickest line). $E^{max}_{nn}(t)$ is the maximum value of the magnitude of the electric field between nn sites, and lightning occurs within the cloud (air) when $E^{max}_{nn}(t)$ exceeds the threshold for conduction in the cloud (air). Perpendicular axes are shifted, and units are the same.
**b.** $E^{max}_{nn}(t)$ for $t_{wait}$ = 100 (N=10), where the number of sites interchanged is varied from zero (thinnest line) to 500 (=0.50x10x10x10) (thickest line). $E^{max}_{nn}(t)$ and perpendicular axes are same as in Fig. 2. a.
**c.** $E^{max}_{nn}(t)$ for $t_{wait}$ = 10 (N=10), where the number of sites interchanged is varied from zero (thinnest line) to 500 (=0.50x10x10x10) (thickest line). $E^{max}_{nn}(t)$ and perpendicular axes are same as in Fig. 2. a.

**Figure 3.**     **a.** $E^{max}_{nn}(t)$ for $t_{wait}$ = 100 (N=5, and pressure increases by a factor of eight at maximum), where the number of sites interchanged is varied from zero (lowest plot) to 62 (=0.50x5x5x5) (highest plot). $E^{max}_{nn}(t)$ is the maximum value of the magnitude of the electric field between nn sites, and lightning occurs within the cloud (air) when $E^{max}_{nn}(t)$ exceeds the threshold for conduction in the cloud (air). Perpendicular axes are shifted, and units are the same.

**b.** $E^{max}_{nn}(t)$ for $t_{wait} = 10$ (N=5, and pressure increases by a factor of eight at maximum), where the number of sites interchanged varied from zero (lowest plot) to 62 (=0.50x5x5x5) (highest plot). $E^{max}_{nn}(t)$ and perpendicular axes are same as in Fig. 3. a.

**Figure 4.**   Lightning towards the earth, where we have bipolar distribution of charges, i.e. I=2.

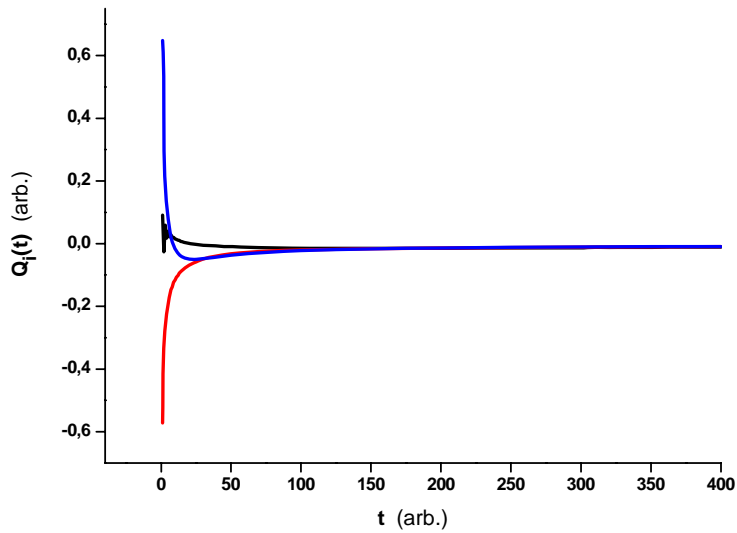

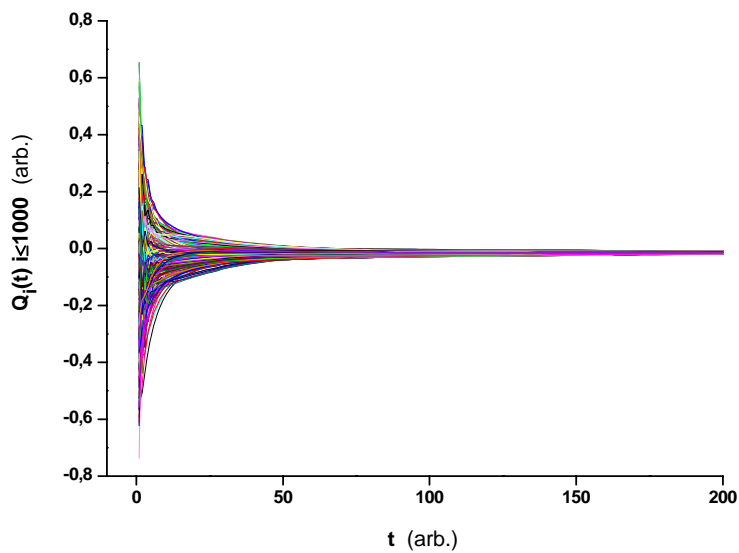

**Figure 1. a.**

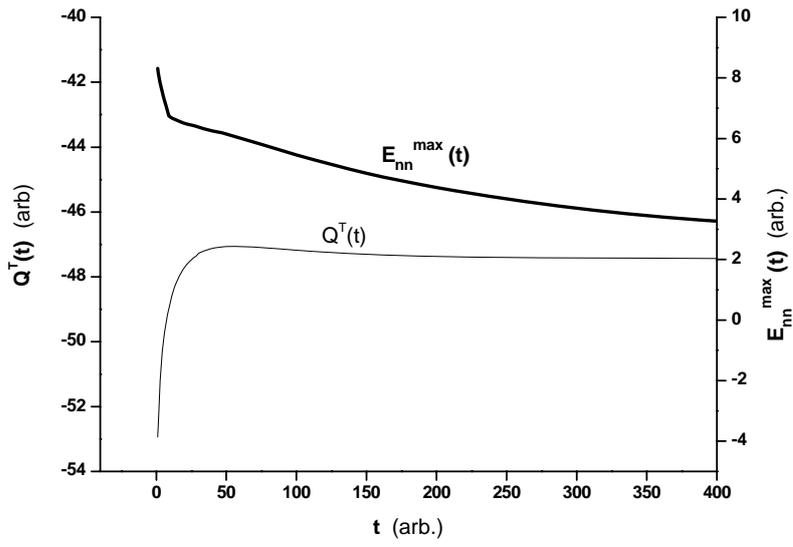

**Figure 1. b.**

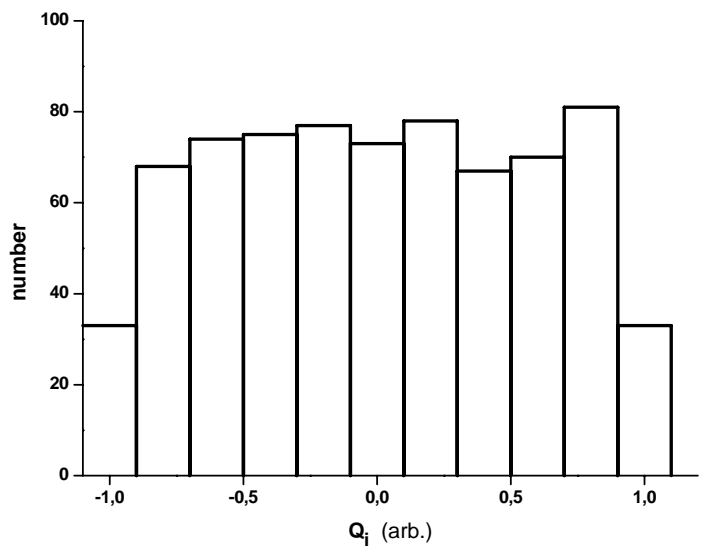

**Figure 1. c.**

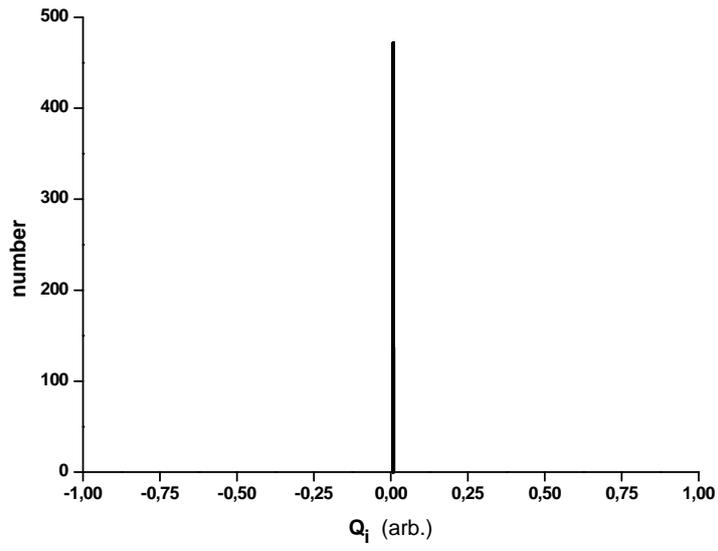

**Figure 1. d.**

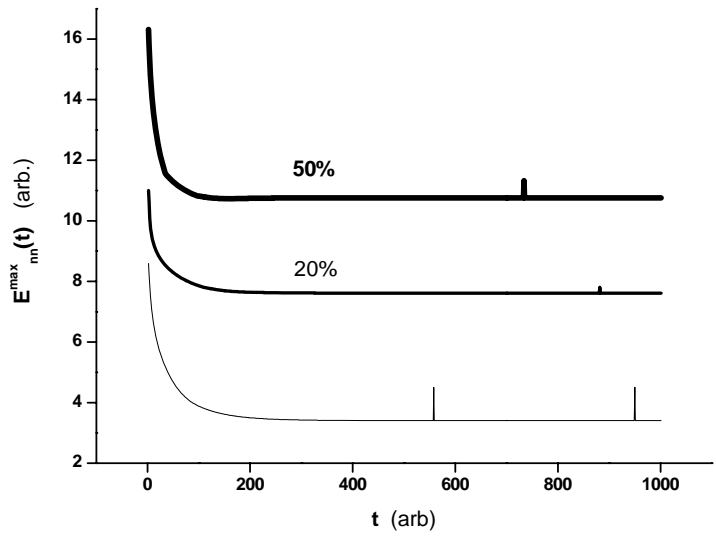

**Figure 2. a.**

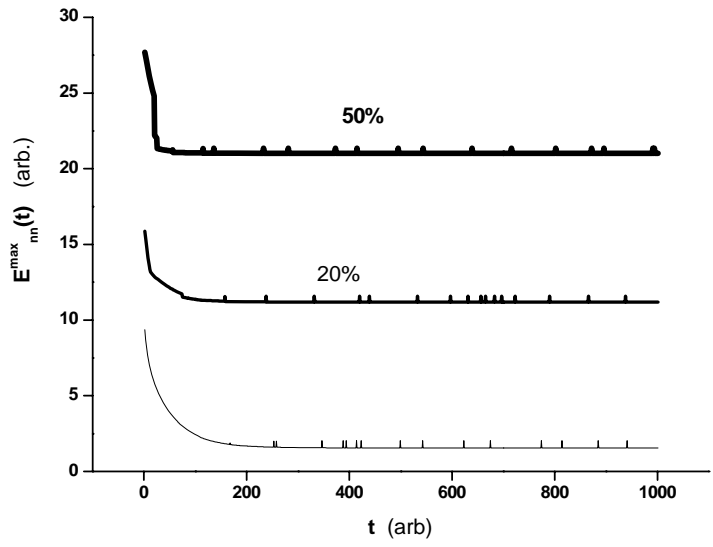

**Figure 2. b.**

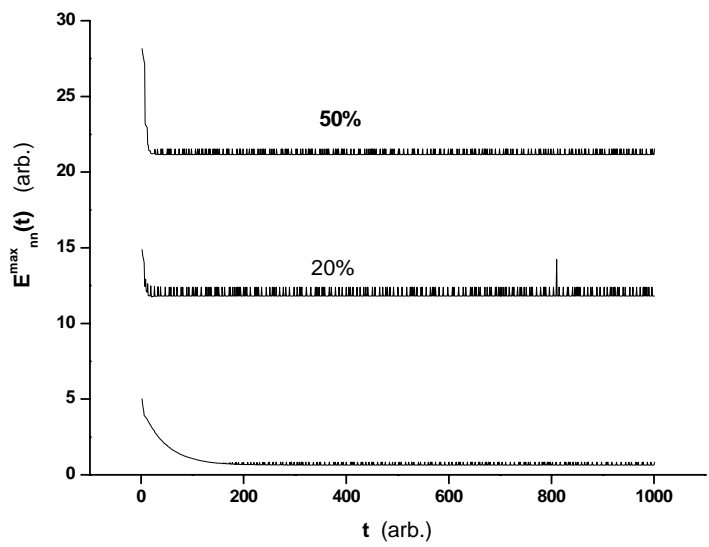

**Figure 2. c.**

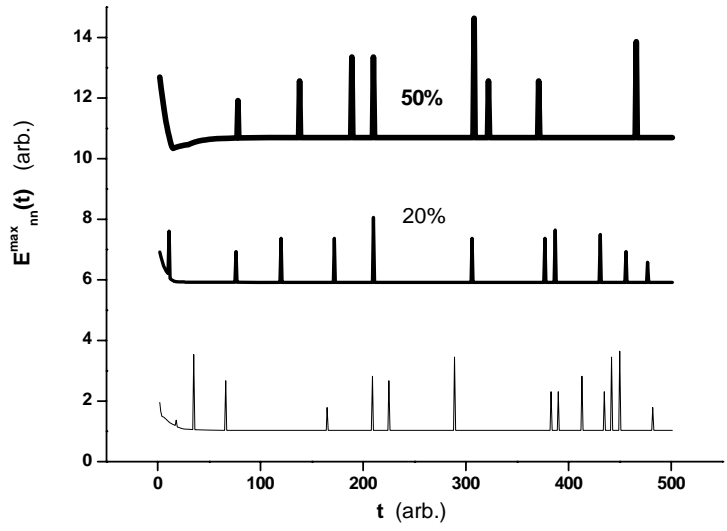

**Figure 3. a.**

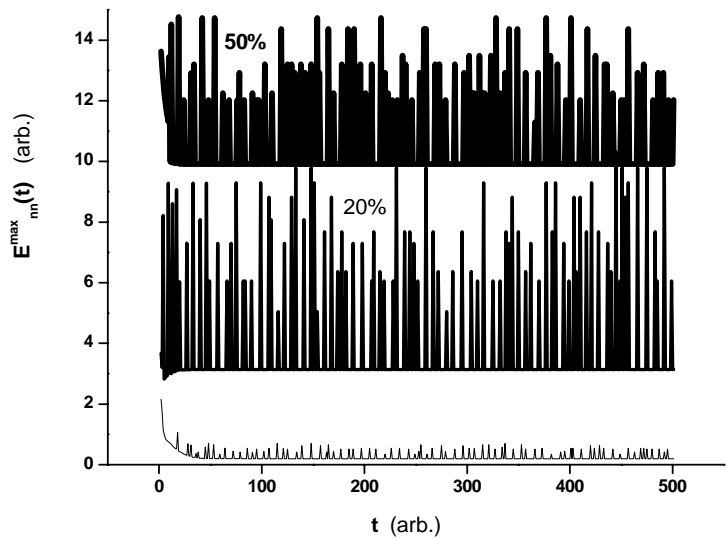

**Figure 3. b.**

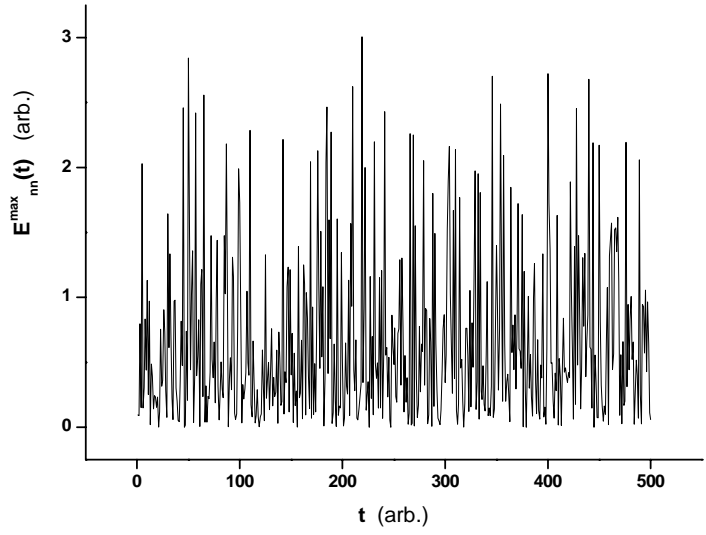

**Figure 4.**